\documentclass[aps,prd,twocolumn,superscriptaddress,nofootinbib]{revtex4}
\usepackage{amsmath,amsfonts,amssymb}
\usepackage{calligra}
\usepackage{color}
\usepackage{units}
\usepackage{graphicx}

\DeclareMathAlphabet{\mathcalligra}{T1}{calligra}{m}{n}
\newcommand{\rth}{\mathrm{th}}
\newcommand{\hden}{\mathcalligra{h}\,\,}
\newcommand{\el}{\mathcalligra{l}\,\,\,}

\usepackage{color}
\definecolor{orange}{rgb}{0.9,0.5,0}

\begin{document}

\title{Rotating neutron stars with non-barotropic thermal profile}

\author{Giovanni~Camelio}
\affiliation{Department of Astronomy and The Oskar Klein Centre, Stockholm University, AlbaNova, 10691 Stockholm, Sweden}
\author{Tim~Dietrich}
\affiliation{Nikhef, Science Park, 1098 XG Amsterdam, The Netherlands}
\author{Miguel~Marques}
\noaffiliation
\author{Stephan~Rosswog}
\affiliation{Department of Astronomy and The Oskar Klein Centre, Stockholm University, AlbaNova, 10691 Stockholm, Sweden}
\date{\today}

\begin{abstract}
Neutron stars provide an excellent laboratory for physics under the most extreme
conditions. Up to now, models of axisymmetric, stationary, differentially
rotating neutron stars were constructed under the strong assumption of barotropicity,
where a one-to-one relation between all thermodynamic quantities exists.
This implies that the specific angular momentum of a matter element depends
only on its angular velocity. The physical conditions in the early stages of
neutron stars, however, are determined by their violent birth processes,
typically a supernova or in some cases the merger of two neutron stars,
and detailed numerical models show that the resulting stars are by no means barotropic.
Here, we construct models for stationary, differentially rotating, non-barotropic
neutron stars, where the equation of state and the specific angular momentum
depend on more than one independent variable. We show that the potential
formulation of the relativistic Euler equation
can be extended to the non-barotropic case, which, to the best of our knowledge,
is a new result even for the Newtonian case. We implement the new method
into the XNS code and construct equilibrium configurations for non-barotropic
equations of state. We scrutinize the resulting configurations by evolving
them dynamically with the numerical relativity code BAM, thereby
demonstrating that the new method indeed produces stationary,
differentially rotating, non-barotropic neutron star configurations.
\end{abstract}

\keywords{}

\maketitle

\section{Introduction}

Black holes and neutron stars are the final stages of the evolution of massive
stars, and they are typically born in supernova explosions or, less frequently,
in binary neutron star mergers. Neutron stars are of particular interest since
they allow for the study of matter properties under extreme density and
temperature conditions that cannot be reached in any terrestrial laboratory,
e.g., \cite{Lattimer:2012nd, Lattimer:2015nhk, Abbott:2018exr, Perego:2019adq,
Fischer+2011}.  These matter properties, however, leave an imprint in the
post-merger gravitational wave signal (at kHz frequencies) that will be
accessible to ground-based gravitational wave detectors of the next generation, 
e.g, \cite{Punturo:2010zz,Sathyaprakash:2019yqt}. 
Moreover, these properties impact also the post-merger
neutrino and electromagnetic signals \cite{Abbot+2017PRL, Abbot+2017ApJL,
Coulter+2017, Burrows1988}.

Stationary rotating equilibrium configurations are often used as
idealizations of the post-merger remnant or as initial conditions
for long-term evolutions  and explorations of the parameter space
\cite[e.g.,][]{Bonazzola+1993, Goussard+1997,
Bucciantini+Del_Zanna2011, Pili+2014, Camelio+2018}.
Thermal effects are in such studies included by
assuming that all thermodynamical quantities, including the
temperature, are functions of only one independent variable,
e.g.\@ the pressure. This leads to ``effective barotropic'' or simply
``barotropic'' stellar models which are particularly convenient
because they allow to write the Euler equation as a potential.

The barotropic assumption is also commonly used to model Newtonian
(e.g., main sequence) stars. In the context of Newtonian stars, however,
non-barotropic stellar models (also called ``baroclinic'') have been computed
both perturbatively \cite{Roxburgh+Strittmatter1966, Clement1969, Monaghan1971,
Sharp+1977} and non-perturbatively \cite{Uryu+Eriguchi1994, Roxburgh2006, Espinosa_Lara+Rieutord2007,
Espinosa_Lara+Rieutord2013, Yasutake+2015, Fujisawa2015}, and even
for Newtonian accretion disks with an analytic procedure \cite{Amendt+1989, Razdoburdin2017}.
In a non-barotropic star, the thermodynamical quantities depend on
more than one independent variable, for example on the pressure and
the temperature, and the Euler equation needs to be solved numerically.
While baroclinic stationary stars are known and studied in Newtonian theory,
they have not yet been addressed in a General Relativity
context\footnote{\citet{Bardeen1970} explicitly considers a
general entropy distribution in the formulation of his variational principle,
but does not compute any stellar structure.}. This is
probably due by the difficulty of solving the Euler equation in differential
form and the fact that thermal effects influence the neutron star structure
only for the first few tens of seconds and are negligible thereafter.

Nevertheless, since  post-merger and post-supernova remnants
are not barotropic \cite[e.g.,][]{Perego:2019adq,Fischer+2011}, or,
more generally, since the lack of non-barotropic models in General Relativity
represents a serious gap in the theory of stellar structure, we want to address this
topic here. We address the non-barotropicity of relativistic neutron stars,
both theoretically and with stationary and dynamical numerical codes.
The novelty of our work is twofold: on the one hand this is the first study
in General Relativity of stationary, differentially rotating, non-barotropic stars;
on the other hand we demonstrate that also in the non-barotropic case
the Euler equation can be cast in the form of a potential. The latter result is
novel even in the Newtonian context.

The paper is organized as follow. We discuss in Sec.~\ref{sec:literature} how thermal effects
are commonly included in barotropic neutron star models. Sec.~\ref{sec:non-barotropic}
describes our novel approach and its numerical implementation is explained
in Sec.~\ref{sec:numeric}. The new approach is validated in Sec.~\ref{sec:results} and Sec.~\ref{sec:discussion}  discusses
some of its implications. We finally summarize and conclude in Sec.~\ref{sec:conclusions}. In three
appendices we describe the Newtonian limit of the (relativistic) Euler equation
(Appendix~\ref{app:newton}) and the non-barotropic (Appendix~\ref{app:eos}) and the effective barotropic
(Appendix~\ref{app:baro}) equations of state adopted.

\section{Rotating stars in General Relativity}
\label{sec:literature}

Unless stated otherwise, we use $\mathrm c=\mathrm G=\mathrm M_\odot=k_\mathrm{B}=1$,
which are also our code units.
Useful conversions to this unit system are
$\unit{km}\simeq0.677$,
$\unit{ms}\simeq203$, and
$\rho_n\simeq4.34\times10^{-4}$,
where $\rho_n$ is the nuclear saturation rest mass density ($\rho_n\simeq\unit[2.68\times10^{14}]{g/cm^3}$).

In this work we are interested in solutions of stationary rotating stars in General Relativity.
We will assume axisymmetry, since non-axisymmetric rotating
bodies radiate gravitational waves and therefore are not stationary.
We will further assume a circular spacetime, which implies the assumption
that meridional currents and convection are negligible.
Under these assumptions, the spacetime shaped by the rotating neutron star
in quasi-isotropic coordinates reads~\cite{Stergioulas2003}:
\begin{multline}
\label{eq:quasi-isotropic-metric}
\mathrm d\tau^2 = -\alpha^2\mathrm dt^2 + A^2(\mathrm dr^2 + r^2 \mathrm d\theta^2)\\
+ B^2r^2\sin^2\theta(\mathrm d\phi - \omega\mathrm dt)^2,
\end{multline}
where $\tau$ is the proper time, $t,r,\theta,\phi$ are the coordinate time, radius,
polar angle, and azimuth angle, respectively, and $\alpha,A,B,\omega$ are metric fields
that depend only on $r,\theta$ due to the stationarity and axisymmetry condition.
$\alpha$ is the lapse and $\omega$ is the angular velocity of the zero angular
momentum observer (ZAMO) as measured by an observer at infinity \cite{Bardeen1970}.
It is useful to define the cylindrical radius (which in General Relativity
has not cylindrical isosurfaces):
\begin{equation}
R(r,\theta) = B(r,\theta)r\sin\theta.
\end{equation}

With these assumptions, the Einstein equations reduce to four equations for the
metric fields $\alpha, A,B,\omega$.  Let us assume that the stellar matter is
described by a perfect fluid,
with energy-momentum tensor
\begin{equation}
\label{eq:tmunu}
T^{\mu\nu}= \hden u^\mu u^\nu + p g^{\mu\nu},
\end{equation}
where $u^\mu$ is the 4-velocity, $p$ is the pressure, and $\hden$ is the total
enthalpy per volume. The Euler equation can be derived from the vanishing
of the covariant divergence of the energy-momentum tensor as
\begin{equation}
\label{eq:iomes}
\frac{\partial_i p}{\hden} + \partial_i{\ln\frac \alpha \gamma}
+F \partial_i \Omega =0,
\end{equation}
where $i = \{r,\theta\}$ [see Appendix~\ref{app:newton}
for the Newtonian limit of Eq.~\eqref{eq:iomes}].
$\gamma$ and $\Omega$ are respectively the Lorentz factor with respect to the ZAMO
and the matter angular speed seen at infinity,
\begin{align}
\label{eq:lorentz}
\gamma={}& \frac{1}{\sqrt{1 - (Rv^\phi)^2}},\\
\label{eq:vphi}
\Omega ={}& \alpha v^\phi + \omega,
\end{align}
where $v^\phi$ is the contravariant matter 3-velocity with respect to the ZAMO,
and $F$ is:
\begin{equation}
\label{eq:fdef}
F = u^t u_\phi = \frac{R^2(\Omega-\omega)}{\alpha^2 - R^2(\Omega-\omega)^2}.
\end{equation}

The specific (per unit energy) angular momentum of a fluid element is given by
\begin{equation}
\label{eq:l}
\el=-\frac{u_\phi}{u_t}=\frac{R^2(\Omega-\omega)}{\alpha^2 + R^2\omega(\Omega - \omega)},
\end{equation}
which is equivalent to
\begin{equation}
\label{eq:fandl}
F=\frac{\el}{1-\Omega\el}.
\end{equation}
Since for axisymmetry and stationarity $F=F(r,\theta)$, it follows that in
general $\Omega=\Omega(r,\theta)$ and $\el=\el(r,\theta)$.

Stationary numerical solutions of the structure of relativistic rotating stars
can be obtained by iteratively solving the metric and matter equations~\cite{Stergioulas2003}.
In the following sections, we will discuss the equations for matter fields.
This
means in particular that the metric fields $\alpha,A,B,\omega$ are known and
fixed from the previous iteration.

\subsection{Isentropic EOS and rigid rotation}
\label{ssec:cr}

Considering an equation of state (EOS) depending on two variables with a thermal part, 
the first law of thermodynamics for the specific enthalpy reads
\begin{equation}
\label{eq:h1}
\mathrm dh = \frac {\mathrm dp}{\rho} + \frac{T}{m_n}\mathrm ds,
\end{equation}
where $\rho$ is the rest-mass density,
$h$ the specific total enthalpy ($h=\hden/\rho$), $T$ is the temperature, $m_n$
the nucleon mass, and $s$ the entropy per baryon.
Since one can get $\rho$ and $T$ from partial differentiation of $h$ with
respect to $p$ and $s$,
\begin{align}
\label{eq:p_maxwell}
\frac1\rho ={}& \left.\frac{\partial h}{\partial p}\right|_s,\\
\label{eq:t_maxwell}
T={}& m_n\left.\frac{\partial h}{\partial s}\right|_p,
\end{align}
it is natural to use the pair $p,s$ as independent variables for the enthalpy
and its derived quantities,
\begin{equation}
\label{eq:h2}
\mathrm dh(p,s) = \frac {\mathrm dp}{\rho(p,s)} + \frac{T(p,s)}{m_n}\mathrm ds.
\end{equation}
If the entropy is uniform in the star, then $\mathrm ds=0$ and\footnote{For
simplicity we use in this work the same symbol for functions that represent the
same physical quantity but depend on different independent variables, even if
mathematically they differ since they are defined on different domains.  We
will always specify the independent variables if they are not clear from the
context.} $h=h(p)$, namely
the EOS is barotropic (i.e., 1D), and the first law of
thermodynamics reads
\begin{equation}
\label{eq:h3}
\mathrm d\ln h = \frac{\mathrm dp}{\hden}.
\end{equation}

In rigid rotation $\partial_i\Omega=0$, and thanks to Eq.~\eqref{eq:h3},
we can write Eq.~\eqref{eq:iomes} as
\begin{equation}
\partial_i\ln h + \partial_i \ln\frac\alpha \gamma = 0,
\end{equation}
which is equivalent to
\begin{equation}
\label{eq:IOMEs_cold_rigid}
\ln h(p) + \ln\frac{\alpha}{\gamma} = \mbox{const},
\end{equation}
where we can determine the constant from  the known central values
of the enthalpy $h_0$ and the lapse $\alpha_0$ (on the axis $Rv^\phi=0$ and therefore $\gamma=1$):
\begin{equation}
\label{eq:const_cold_rigid}
\mathrm{const}=\ln(h_0\alpha_0).
\end{equation}
From Eqs.~\eqref{eq:IOMEs_cold_rigid}--\eqref{eq:const_cold_rigid} and fixing the uniform
angular velocity $\Omega=\Omega_0$ one can easily get $h$ and from it $p$
and the other EOS quantities.

The most common example of neutron stars studied in the literature
are cold stars (i.e., uniform vanishing entropy per baryon).
An example of cold, rigidly rotating neutron star is marked as ``CR''
in this paper.

\subsection{Barotropic EOS and differential rotation}
\label{ssec:baro-diff}

Under the assumption that the entropy per baryon depends only on the pressure
$s=\tilde s(p)$, a hot EOS depends on pressure alone, i.e.\@ it becomes an
effective barotrope:
\begin{equation}
h(p)=h\big(p,\tilde s(p)\big).
\end{equation}
This can be observed in Fig.~\ref{fig:s}, where we show the entropy per baryon as a
function of the rest-mass density in the interior of a neutron star. The black
lines correspond to the effective barotropic assumption, while the red regions
are obtained by dropping this assumption as described in Sec.~\ref{sec:non-barotropic}.
It is convenient to define the ``heat function''
\begin{equation}
\label{eq:heatdef}
H(p) = \int_{p_0}^{p}\frac{\mathrm dp'}{\hden(p')},
\end{equation}
where $p_0$ is the given central pressure,
from which we obtain
\begin{equation}
\partial_i H(p)= \frac{\partial_i p}{\hden}.
\end{equation}

Additionally, if we assume that $F$ depends only on $\Omega$, we have analogously:
\begin{align}
\label{eq:fintegral}
\mathcal F(\Omega) ={}& \int_{\Omega_0}^{\Omega} F(\Omega')\mathrm d\Omega',\\
\label{eq:fderiv}
\partial_i \mathcal F(\Omega)={}& F(\Omega)\partial_i \Omega,
\end{align}
where $\Omega_0$ is the given angular frequency on the symmetry axis
and $\mathcal F(\Omega)$ is called ``differential-rotation law''.
Using Eqs.~\eqref{eq:heatdef}-\eqref{eq:fderiv},
Eq.~\eqref{eq:iomes} is equivalent to
\begin{equation}
\label{eq:IOMEs_hot_differential}
H(p) + \ln\frac\alpha \gamma + \mathcal F(\Omega) =  \ln \alpha_0.
\end{equation}
One can determine the matter properties in every point $(r,\theta)$
by determining $\Omega$ from the relation $\mathcal F'(\Omega)=F(\Omega,r,\theta)$,
where we show explicitly the dependence on the yet-to-be-determined $\Omega$,
and then $p$ from Eqs.~\eqref{eq:heatdef} and
\eqref{eq:IOMEs_hot_differential}. The other EOS quantities are easily
determined because the EOS is effectively barotropic.

For an isentropic star it is $H(p)=\ln h(p) - \ln h_0$, and if in addition the star
is in rigid rotation,
one recovers Eq.~\eqref{eq:h3}, as expected.

One can assume an analytic form for the differential-rotation law, for example
by adopting the ``j-const'' law that is commonly used in literature (\cite{Komatsu+1989a},
see also \cite{Uryu+2017, Witzany+Jefremov2018}):
\begin{equation}
\label{eq:f_law}
\mathcal F(\Omega)= -\frac{R_0^2}2(\Omega - \Omega_0)^2,
\end{equation}
where $R_0$ has the dimension of a length and sets the scale of the
differential rotation, that is, $\Omega\simeq\Omega_0/2$ at $R=R_0$
\cite{Villain+2004}. Rigid rotation cannot be described by a
differential-rotation law because $\Omega$ is constant, but $F$ is not. Therefore,
it can only be recovered in the limit $R_0\to\infty$.
To model rigid rotation, one can just fix $\Omega=\Omega_0$ and drop the
$\mathcal F$ term in Eq.~\eqref{eq:IOMEs_hot_differential}; however in Sec.~\ref{ssec:leg}
we show how it is possible to cleanly unify the description of rigidly and differentially
rotating stars.

The assumption $F=F(\Omega)$ is equivalent to requiring that $\el=\el(\Omega)$
[cf.~Eq.~\eqref{eq:fandl}], namely it is equivalent
to dropping any dependence on the metric and the coordinates in the relation between the
specific angular momentum and the angular speed. This can be seen in Fig~\ref{fig:el},
where we show the specific angular momentum as a function of the angular velocity
in the interior of a neutron star. The black line corresponds to the case discussed
in this section, where the specific angular momentum is in a one-to-one correspondence
with the angular velocity, while the red region
is obtained by dropping this assumption as described in Sec.~\ref{sec:non-barotropic}.

\section{Non-barotropic thermal profile}
\label{sec:non-barotropic}

\begin{figure}
\includegraphics[width=\columnwidth]{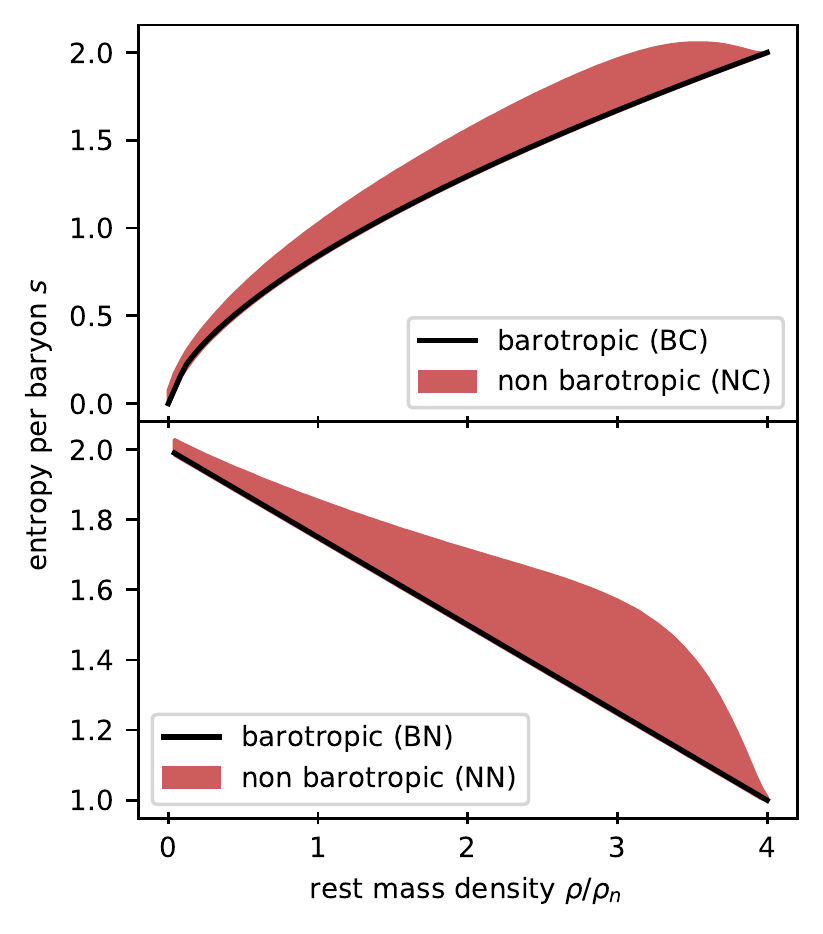}
\caption{\label{fig:s}Entropy per baryon $s$ as a function of rest-mass density $\rho$
for 2 barotropic (black lines) and 2 non-barotropic (red regions) models considered
in this paper, cf.~Table~\ref{tab:abb}.
The upper/lower edge of the red regions
corresponds to the entropy along the
equatorial plane/rotational axis of the non-barotropic neutron star, respectively.
Similar plots obtained from dynamical simulations are e.g.\@ Fig.~1 of
\citet{Fischer+2011} and Figs.~3--8 of
\citet{Perego:2019adq}.}
\end{figure}

\begin{figure}
\includegraphics[width=\columnwidth]{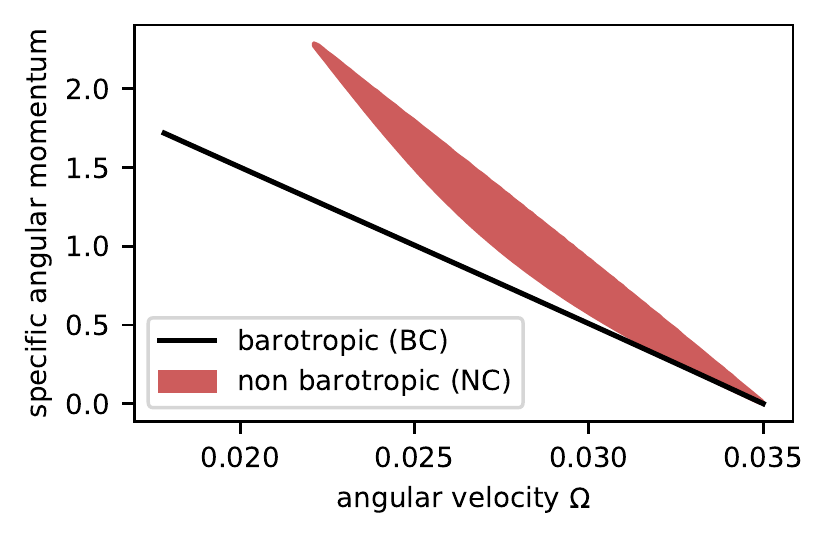}
\caption{\label{fig:el}Angular momentum per unit energy $\el$ as a function of
angular velocity $\Omega$ for a barotropic (black line) and a non-barotropic
(red region) model considered in this paper, cf.~Tab.~\ref{tab:abb}. The upper/lower
edge of the red region corresponds to the specific angular momentum along the
equatorial plane/stellar border, respectively.  Non-convective models behave
similarly.}
\end{figure}

The big problem of the method described in the previous section is that
one is limited to an effective barotropic EOS, i.e.\@ the EOS
is actually a function of one independent variable only, even in presence of thermal effects.
Similarly, one enforces $\el=\el(\Omega)$, dropping any dependence on
the metric, see black lines in
Figs.~\ref{fig:s} and \ref{fig:el}.
However, dynamical core-collapse supernova and binary neutron star merger simulations
show that realistic newly-born neutron stars are non-barotropic \cite[e.g.,][]{Fischer+2011,
Perego:2019adq}.

In this section we show how it is possible to overcome these limitations in
a rigorous way.

\subsection{The generalization}

\begin{figure}
\includegraphics[width=\columnwidth]{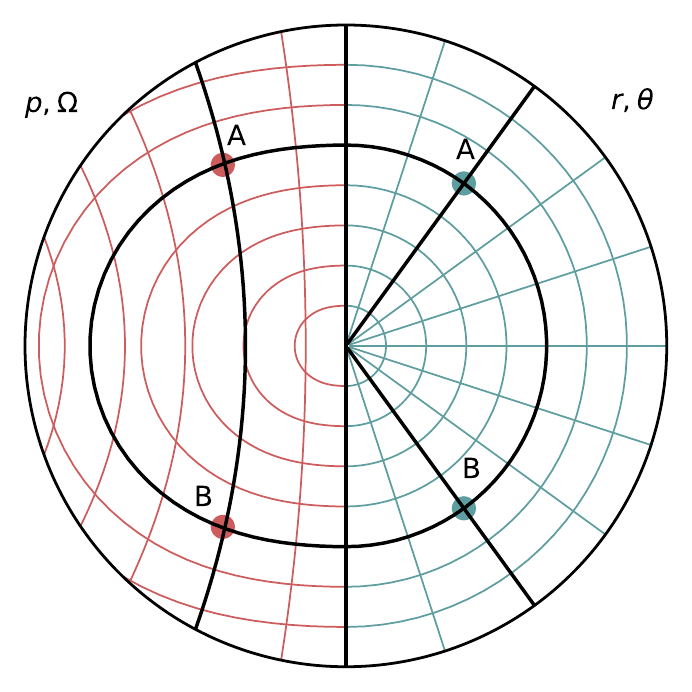}
\caption{\label{fig:coord}Sketch of the coordinate grid in $p,\Omega$ (left,
red) and in $r,\theta$ (right, blue).  The $p$ coordinate is elliptic-like
while the $\Omega$ coordinate is parabolic-like, cf.~Fig.~\ref{fig:xns}.
Note that the planar symmetric A and B points have different $\theta$
coordinate but the same $p,\Omega$ coordinates.}
\end{figure}

Eq.~\eqref{eq:iomes} can be written as
\begin{equation}
\label{eq:IOMEs_standard}
\frac{\mathrm dp}{\hden} + F\mathrm d\Omega + \mathrm d \ln\frac\alpha \gamma = 0,
\end{equation}
to stress that when $\hden= \hden(p)$ and $F=F(\Omega)$ it is
\begin{equation}
\mathrm d\left( H(p) + \mathcal F(\Omega) + \ln\frac\alpha \gamma\right)=0,
\end{equation}
namely the Euler equation implies the existence of a conserved quantity and
\begin{align}
\label{eq:rhoh_maxwell}
\frac 1{\hden} ={}& \frac{\mathrm d H(p)}{\mathrm d p},\\
\label{eq:F_maxwell}
F ={}& \frac{\mathrm d\mathcal F(\Omega)}{\mathrm d \Omega}.
\end{align}

In other words, we are casting the Euler equation in a
potential form similar to Thermodynamics.
However, comparing the
thermodynamical case [e.g., Eqs.~\eqref{eq:p_maxwell}--\eqref{eq:t_maxwell}] with the stellar case
[i.e., Eqs.~\eqref{eq:rhoh_maxwell}--\eqref{eq:F_maxwell}], one notes that in contrast to the former, in
the latter we are determining the derived quantities with total derivatives of two potentials
instead of partial derivatives of one potential.
\emph{Here we push the similarity with Thermodynamics one
step further.}

Let us pursue this intuition:
\begin{align}
\label{eq:i_lnaw}
Q(p,\Omega)={}& -\ln\frac\alpha \gamma,\\
\label{eq:new_iomes}
\partial_i Q(p,\Omega)={}& \frac{\partial_ip}{\hden(p,\Omega)} + F(p,\Omega)\partial_i\Omega,\\
\label{eq:oneoverhden}
\frac{1}{\hden(p,\Omega)} ={}& \left.\frac{\partial Q(p,\Omega)}{\partial p}\right|_\Omega,\\
F(p,\Omega) ={}& \left.\frac{\partial Q(p,\Omega)}{\partial \Omega}\right|_p,
\end{align}
where we defined the potential $Q$ and
all quantities depend on $p,\Omega$ because these are the natural variables for
the same reason $p$ and $s$ are the natural variables for the thermodynamical case, namely because the
other quantities ($\hden$ and $F$ in the stellar case, $\rho$ and
$T$ in the thermodynamical case) can be determined from partial differentiation
with respect to those. 
Note that Eq.~\eqref{eq:new_iomes} is exactly the Euler equation [Eq.~\eqref{eq:iomes}]
and that it mirrors the equivalent thermodynamical equation [after substituting the exact
differential with partial differentiation in Eq.~\eqref{eq:h2}].

We should be careful because for axisymmetry and stationarity it is also
$Q=Q(r,\theta)$, $p=p(r,\theta)$, and $F=F(r,\theta)$: given
the pair $p$ and $\Omega$, we must be able to determine the pair $r$ and $\theta$.
However, this change of coordinates is not bijective, that is, each pair
$p$ and $\Omega$ corresponds to two pairs $r$ and $\theta$, one in the northern hemisphere
and one in the southern hemisphere, and therefore to two potentials: $Q_+(p,\Omega)$
and $Q_-(p,\Omega)$, that are identical in the planar case $Q_+=Q-$.
In Fig.~\ref{fig:coord} we show how the interior of a star is mapped with
$r$ and $\theta$ coordinates (on the right) and with $p$ and $\Omega$ coordinates (on the left).

The key point here is that the additional dependence of $\hden$ on $\Omega$
[as opposed to a dependence only on $p$, see Eq.~\eqref{eq:oneoverhden}]
``breaks'' the barotropicity because, as can be seen in Fig.~\ref{fig:coord},
$\Omega$ is not in a one-to-one correspondence with $p$.
This additional dependence is made possible by allowing for
$\partial_p\partial_\Omega Q\neq 0$.

It is worth noting that:
\begin{itemize}
\item The standard case described in Sec.~\ref{ssec:baro-diff} is
equivalent to the following potential:
\begin{equation}
\label{eq:q_std}
Q(p,\Omega) = H(p) + \mathcal F(\Omega) -\ln\alpha_0.
\end{equation}
\item Since we rewrote Eq.~\eqref{eq:iomes} in terms of a potential,
the difference of pressure and angular speed between two stellar points
does not depend on the integration path but only on the initial and final points.
\item From the Schwarz's theorem we get the Maxwell-like relation
\begin{equation}
\left.\frac{\partial \hden^{-1}}{\partial \Omega}\right|_p
=
\left.\frac{\partial F}{\partial p}\right|_\Omega.
\end{equation}
\end{itemize}

\subsection{A simple non-barotropic model}

Assuming that the analytic form of $Q(p,\Omega)$ is known, 
but that we do not know the pressure and angular velocity profiles 
$p(r,\theta)$ and $\Omega(r,\theta)$, 
we have to solve the following system of equations
in every point:
\begin{align}
\label{eq:sys1}
Q(p,\Omega) ={}& -\ln\frac{\alpha(r,\theta)}{\gamma(r,\theta,\Omega)},\\
\label{eq:sys2}
\partial_\Omega Q(p,\Omega)={}& F(r,\theta,\Omega),\\
\label{eq:sys3}
\partial_pQ(p,\Omega) ={}& \frac1{\hden\big(p,s(r,\theta)\big)}.
\end{align}
In Eqs.~\eqref{eq:sys1}-\eqref{eq:sys3} we have made explicit the dependence of 
every quantity on the position
in the star $(r,\theta)$ and on the yet-to-be-determined quantities $(p,\Omega)$.
Given a point in the star $(r,\theta)$ and the entropy in that point
$s(r,\theta)$, this is a system of 3 equations in 2 variables ($p,\Omega$), that in general
has no solution.
On the other hand, if we leave $s(r,\theta)$ undetermined, given $(r,\theta)$ we can first
determine $(p,\Omega)$ solving Eqs.~\eqref{eq:sys1}--\eqref{eq:sys2}, and then
determine $s(r,\theta)$ from Eq.~\eqref{eq:sys3}.

Let us now consider a simple\footnote{Note that this is not the only
potential that generalizes the standard case; for example another valid choice
is obtained by substituting $Q_0\to0$ and $H(p)\to
H(p)-\ln\alpha_0$ in Eq.~\eqref{eq:case_c}, which gives a different but still consistent
solution.} non-trivial case:
\begin{equation}
\label{eq:case_c}
Q(p,\Omega) = Q_0 + H(p) + \mathcal F(\Omega) + bH(p) \mathcal F(\Omega),
\end{equation}
where $b$ is a ``barotropic'' parameter and the constant $Q_0$ is
determined from the condition
$Q_0=Q(p_0,\Omega_0)=-\ln\alpha_0$.
The standard case of Eq.~\eqref{eq:q_std} is re-obtained for $b=0$.
$H$ and $\mathcal F$ are formally defined as in Eqs.~\eqref{eq:heatdef} and \eqref{eq:fintegral}, but have not the same
physical meaning. In particular, the arbitrary barotropic function $\tilde s(p)$
that enters in the definition of $H(p)$
does not correspond to a physical entropy unless $b=0$ (this is the reason we defined it with a tilde).

The potential $Q$ in this form is particularly convenient, because we can
factor out the dependence on $p$ and therefore we have to solve only
one equation to determine $\Omega$. In fact, Eq.~\eqref{eq:sys2} reads
\begin{equation}
\mathcal F'(\Omega)\big(1 + b H(p)\big) = F(r,\theta,\Omega),
\end{equation}
and using the definition~\eqref{eq:case_c} we get
\begin{multline}
\label{eq:root_omega}
\mathcal F'(\Omega)\left(1 + bQ(r,\theta,\Omega)
- bQ_0\right)\\
= F(r,\theta,\Omega)\big(1 + b\mathcal F(\Omega)\big),
\end{multline}
that can be solved for $\Omega$ with a 1D root finding
[$Q(r,\theta,\Omega)$ is the RHS of Eq.~\eqref{eq:sys1}].
Knowing $\Omega$, one can first determine $H(p)$ and then $\hden$ from
\begin{align}
\label{eq:heat_from_p}
H(p)={}& \frac{Q(r,\theta,\Omega)-Q_0 - \mathcal F(\Omega)}{1+b\mathcal F(\Omega)},\\
\label{eq:hden}
\hden(p,\Omega)={}& \frac 1{H'(p) \big(1 + b\mathcal F(\Omega)\big)},
\end{align}
where
$H'(p)$ is the total derivative of $H(p)$.
Knowing $\hden$ and $p$ [obtained from the inversion of $H(p)$] one can use
them to invert the EOS, that in the case considered here depends on two independent
variables (we discuss in Sec.~\ref{ssec:eos3d} how to generalize the procedure
to an EOS that depends on more than two independent variables).

It is useful at this point to recap what we have accomplished.
We have first defined in Eq.~\eqref{eq:case_c} a function $Q(p,\Omega)$ and then enforced
with Eqs.~\eqref{eq:sys1}--\eqref{eq:sys3} that this function acts as a potential for the Euler equation.
In this way both the matter and the rotational profiles of the star
are uniquely determined from the potential $Q$
and are function in general of more than one independent variable, therefore breaking the stellar barotropicity.
In Sec.~\ref{ssec:s} we show how, in principle, one can use the freedom in the
definition of $Q$ to tune the thermodynamical and rotational profiles.

Note that for the non-barotropic models in Figs.~\ref{fig:s} and \ref{fig:el}
(red filled contours) the relations $s=s(\rho)$ and $\el=\el(\Omega)$ do not hold anymore.

\section{Numerical implementation}
\label{sec:numeric}

\subsection{XNS code}
\label{ssec:xns}

The XNSv2 code \cite{Bucciantini+Del_Zanna2011, Pili+2014} determines the stationary structure
of a rotating neutron star in the eXtended Conformal Flatness Condition (XCFC)
approximation \cite{Cordero-Carrion+2009}.
The metric equations are solved with a spherical
harmonics decomposition on the angular direction and with finite
differences along the radial direction.
In the XCFC approximation the
metric equations are simpler and hierarchically decoupled; this approximation
is equivalent to enforce in Eq.~\eqref{eq:quasi-isotropic-metric}
\begin{equation}
\label{eq:XCFC}
A(r,\theta)\equiv B(r,\theta)\equiv \psi^2(r,\theta),
\end{equation}
where $\psi$ is called conformal factor, and it is justified because the maximal
relative difference between the $A$ and $B$ metric functions is of the order of
$10^{-3}$ \cite{Gourgoulhon2010}. 
The XCFC approximation yields results of excellent accuracy
for rotating neutron stars \cite[e.g.,][]{Camelio+2018}, while has been showed to degrade
for differentially rotating neutron stars \cite{Iosif+Stergioulas2014}.
Using the diagnostic formula of Eq.~(20) of \citet{Iosif+Stergioulas2014},
we estimate for the configurations studied in this paper a maximal error for
local quantities (e.g., the angular velocity at the equator) within 2\% and a
much smaller error for global quantities (e.g., the gravitational mass).
The estimated error is adequate for a good description of the rotating
neutron star and its spacetime.
In any case, we emphasize that the non-barotropic theory, which we develop in this paper, does
not depend in any way on the use of the XCFC approximation.

In this paper we use our modified version \cite{Camelio+2018} of XNSv2 
and simply refer to it as XNS in
the following. In \citet{Camelio+2018} we described and validated it
against the RNS code \cite{Stergioulas+Friedman1995}
that solves the stationary configuration of rotating neutron stars in general
relativity without
approximations. We refer the reader to \cite{Cordero-Carrion+2009,
Bucciantini+Del_Zanna2011, Pili+2014, Camelio+2018} for the general structure
of XNS and the XCFC equations and just describe the
main modifications with respect to \cite{Camelio+2018}.

To determine the solution of a rotating star, XNS iterates between the solution
of the metric and the matter equations
until convergence.  When the matter quantities ($\hden,p,v^\phi$) are
updated, the metric quantities ($\alpha,\psi,\omega$) are kept fixed,
and vice versa.
To update the matter quantities, the following procedure is repeated
for each grid point $r_i,\theta_j$ (we start from the center, $r_i=r_1$,
and increase $i$ outward):
\begin{enumerate}
\item If the star is rigidly rotating, set $\Omega=\Omega_0$.\\
Otherwise, determine $\Omega$ from Eq.~\eqref{eq:root_omega}.
\item Find $H(p)$ from Eq.~\eqref{eq:heat_from_p}.
\item Find $p$ inverting $H(p)$.
\item If $p<p_s$ ($p_s$ being a fixed value of the surface pressure), go to step 8.
\item If the star is non-barotropic:
\begin{enumerate}
\item Find $\hden$ from Eq.~\eqref{eq:hden}.
\item If the pair $\hden,p$ is not physical (e.g., $\hden \le p$), go to step 8.
\end{enumerate}
\item All independent quantities have been computed. Solve the EOS from $p$
(if barotropic) or $p,\hden$ (if non barotropic). Determine $v^\phi$ from $\Omega$.
\item Go to step 1 with the next $r_i$.
\item The point is outside the surface. Set to zero all matter quantities in $r\ge r_i$
and go to step 1 with $r_i=r_1$ and the next $\theta_j$.
\end{enumerate}

We adopt a rectangular non-evenly spaced grid in $r,\theta$ \cite{Camelio+2018}.
Our radial grid is divided in two regions: the inner part has 2000 evenly spaced points from
$r=0$ excluded to $r=15$ and the outer part has 2000 increasingly spaced points from $r=15$
to $r=1000$. The angular grid ($0<\theta<\pi$) contains 501 points on the 
Legendre knots. We used 50 angular harmonics in the pseudo-spectral expansion and
we consider the result converged when the maximal absolute variation of the rest-mass density
between two iterations is smaller than $10^{-12}$. The surface pressure is set to $p_s=10^{-40}$
in code units ($c=G=M_\odot=1$).

\subsection{BAM code}

\begin{figure}
\includegraphics[width=\columnwidth]{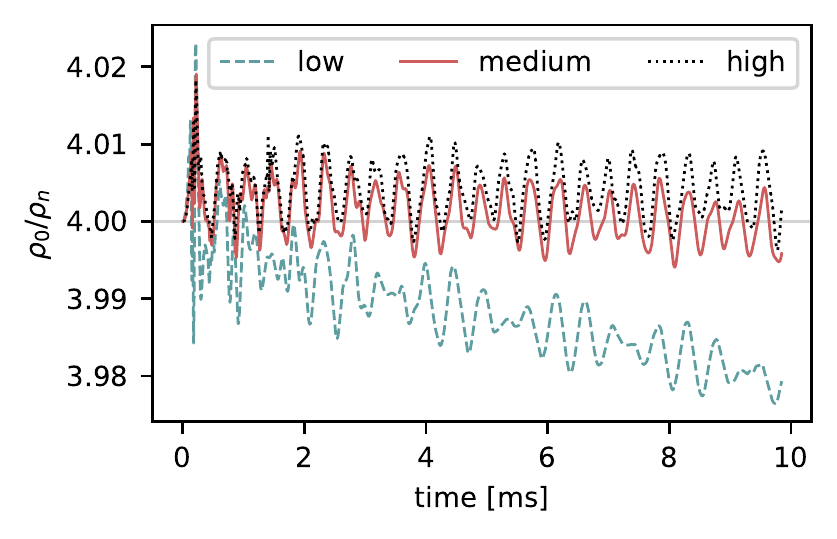}
\caption{\label{fig:res}
BAM evolution of the central rest-mass density of the stellar model CR for different resolutions.}
\end{figure}

We also study the dynamical evolution of the XNS configurations with the 
BAM code~\cite{Bruegmann:2003aw,Brugmann:2008zz,Thierfelder:2011yi,Dietrich:2015iva,
Bernuzzi:2016pie,Dietrich:2018bvi}.
BAM employs a simple mesh refinement scheme where the grid is composed of nested 
Cartesian boxes. 
The grid setup is controlled by the resolution $\Delta x$ in the
finest levels. The outer levels are constructed by progressively coarsening the
resolution by factors of two. We solve the Einstein Equations using the 
Z4c evolution scheme~\cite{Bernuzzi:2009ex,Weyhausen:2011cg,Hilditch:2012fp} and employ fourth order finite-difference stencils. 
The equations of general relativistic hydrodynamics employ a finite-volume 
shock-capturing method and the hydrodynamical flux is computed with the Local Lax-Friedrichs scheme 
using the WENOZ limiter~\cite{Borges:2008a, Bernuzzi:2016pie}. 

The evolution equation system is closed with the EOS, 
for which we assume an ideal gas with a cold and a thermal contribution: 
\begin{equation}
\label{eq:ideal_gas}
p(\rho,u_\rth)= K\rho^\Gamma + (\Gamma_\mathrm{th} - 1)\rho u_\rth,
\end{equation}
where $u_\rth$ is the specific thermal energy and $K,\Gamma,\Gamma_\rth$ are
EOS-dependent parameters, cf.~Appendix~\ref{app:eos} and Tables~\ref{tab:abb} and \ref{tab:models}.

To proof the robustness of our numerical scheme, we show the central rest-mass density evolution of the 
CR model, i.e., of a cold, rigid rotating neutron star, in Fig.~\ref{fig:res}; 
we refer the interested reader to~\cite{Bernuzzi:2011aq,Dietrich:2015iva,
Bernuzzi:2016pie,Dietrich:2018upm,Dietrich:2018phi} for additional tests and convergence analyses. 

We increase the BAM resolution by factors of two, 
where for the low resolution (blue line) the minimum grid resolution in the finest level is $0.1875$, 
the medium resolution (red line) has a minimum grid spacing of $0.09375$, 
and the high resolution (black line) has a minimum grid spacing of $0.046875$. 
This is compatible to the highest resolved binary neutron star 
simulations performed for gravitational wave model development to date~\cite{Dietrich:2019kaq,Kiuchi:2017pte}. 
We save computational costs by simulating only a single quadrant 
of the numerical domain making use of the axisymmetry of the spacetime and
the planar symmetry of the models. 
From Fig.~\ref{fig:res}, we conclude that the changes in the central density decrease 
with increasing resolution. In particular, the central density decrease, which is present in the low resolution case, 
is small for the medium and high resolution. 
The remaining density oscillations of the order of $\sim 0.25\%$ seems 
negligible for the studies discussed in the following\footnote{We remark that the
remaining density oscillations is likely to be related to the XCFC
approximation of XNS, since it is absent or smaller if the XCFC approximation 
is not employed; cf.~Fig.~2 of~\cite{Bernuzzi:2016pie} for single star evolutions 
and the supplementary material of~\cite{Dietrich:2018phi} for studies in binary 
neutron star configurations.}. If not otherwise stated, we will show the results for 
the high resolution grid configuration, but all models have been
simulated with the low, medium, and high grid resolutions to test the correctness of our results. 

\subsection{Models}
\label{ssec:models}

\begin{table}
\begin{tabular}{cl}
\hline
\hline
name & configuration\\
\hline
CR & Cold, Rigidly rotating \\
BC & differentially rotating, Barotropic, Convective \\
NC & differentially rotating, Non-barotropic, Convective \\
C$\Omega$ & Control with $b=0$ in Eq.~\eqref{eq:root_omega} \\
C$p$ & Control with $b=0$ in Eq.~\eqref{eq:hden} \\
BN & differentially rotating, Barotropic, Non-convective \\
NN & differentially rotating, Non-barotropic, Non-convective \\
\hline
\hline
\end{tabular}
\caption{Abbreviated names of the stellar configuration studied in this work.}
\label{tab:abb}
\end{table}

\begin{figure}
\includegraphics[width=\columnwidth]{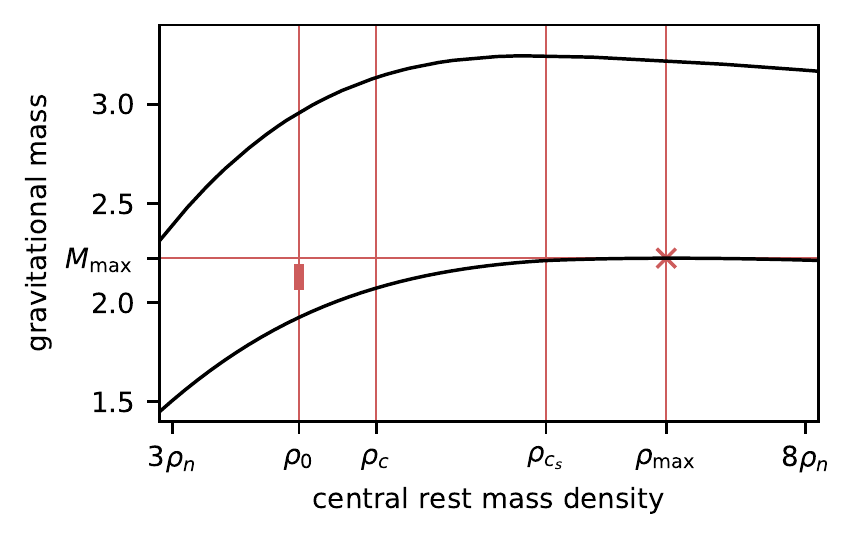}
\caption{\label{fig:mrho}Gravitational mass as a function of the central density for the EOS
adopted in this paper with causality enforced at $\rho>\rho_{c_s}=5.95\rho_n$.
The lower black line corresponds to non-rotating cold models and the upper black line to cold models that rotate rigidly
at the Keplerian limit.
$M_\mathrm{max}=2.22$ is the maximal non-rotating mass corresponding to $\rho_\mathrm{max}=6.90\rho_n$
(red cross)
and $\rho_c=4.60\rho_n$ is the critical density for inverting the non-barotropic EOS (see
Appendix~\ref{app:eos}).
The thick red line marks the region of central density and gravitational (Komar) mass of the models
considered in this paper ($\rho_0=4\rho_n$).}
\end{figure}

To minimize additional code changes in BAM and XNS, we use throughout this work 
an EOS such that the total energy density is given by
\begin{equation}
\epsilon(\rho,s)= \rho + k_1 \rho^\Gamma + k_2s^2 \rho^{\Gamma_\rth},
\end{equation}
where $k_1,k_2,\Gamma,\Gamma_\rth$ are parameters specified in
Table~\ref{tab:models}.  With our parameter choice this EOS has a
maximal cold, non-rotating neutron star mass of $\unit[2.22]{M_\odot}$ as shown in Fig.~\ref{fig:mrho}, and
can be straightforwardly included in BAM, since 
it is equivalent to an ideal gas EOS with $K=(\Gamma - 1)k_1$ (Appendix~\ref{app:eos}).

We fix the barotropic function by setting $\tilde s(\tilde \rho)$.
We remark that with our choice of the potential $Q$, $\tilde \rho$
and $\tilde s$ are physical rest-mass density and entropy per baryon also when $b\neq0$
only on the rotational axis, since there
$\mathcal F(\Omega_0)=0$. For this reason, there is no ambiguity in using the central quantities
in Table~\ref{tab:models}.

\begin{figure*}
\includegraphics[width=\textwidth]{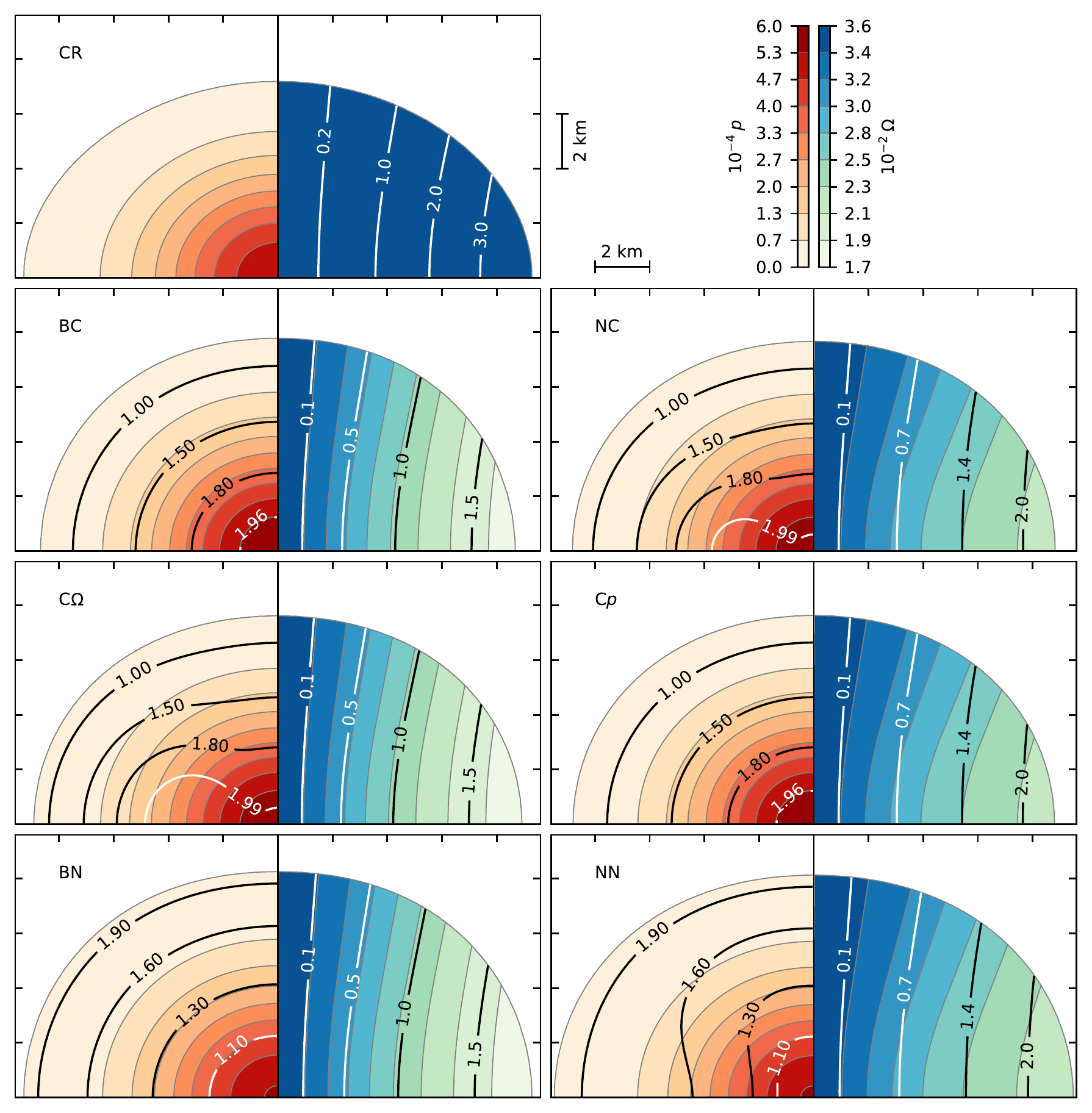}
\caption{\label{fig:xns}Stationary stellar models obtained with XNS.
For each model, the color filled contours refer to the pressure $p$ (red scale, left)
and the angular velocity $\Omega$ (blue scale, right), while the thick black and white contours
to the entropy per baryon $s$ (left)
and the specific angular momentum $\el$ (right).
See text for details.
}
\end{figure*}

We consider 7 models, all shown in Fig.~\ref{fig:xns} and described in Tables~\ref{tab:abb} and \ref{tab:models}.
We remark that if two quantities have parallel level contours means that they are in
a one-to-one correspondence, cf.~Fig.~\ref{fig:xns}.
The control
configurations C$\Omega$ and C$p$ have been obtained with the same procedure as NC, but for
C$\Omega$ we set $b=0$ in Eq.~\eqref{eq:root_omega} and for C$p$ we
set $b=0$ in Eq.~\eqref{eq:hden}. For this reason, $\el=\el(\Omega)$
for C$\Omega$ and $s=s(p)$ for C$p$. Since the potential $Q$ has not been
solved consistently, C$\Omega$ and C$p$ are expected not to be true stationary
solutions and are therefore our control models against which we will judge the
quality of the theory.

The parameters of the EOS and of the potential $Q$ that completely determine the stellar models
are shown in Table~\ref{tab:models}.
The values of parameters $R_0$ and $b$ have been chosen to emphasize differential
rotation and non-barotropicity,
while the choice of the other parameter values is discussed in Appendix~\ref{app:eos}.
All models are stable against dynamical instabilities, i.e., they do not collapse
(Appendix~\ref{app:eos}), 
but some models are unstable against convection (Appendix~\ref{app:baro}).
Note that the obtained central temperatures $T_0$ are reasonable
for proto-neutron stars and for post-merged neutron stars.

More details on the EOS and the rationale behind our choices are provided in
Appendices~\ref{app:eos} and \ref{app:baro}.

\section{Results}
\label{sec:results}

\begin{table*}
\setlength{\tabcolsep}{4pt}
\begin{tabular}{cccccccccccccccc}
\hline
\hline
name & $\Gamma$ & $k_1$ & $K$ & $\Gamma_\rth$ & $k_2$ &
$\rho_0$ & $\tilde s(\tilde\rho)$ & $\Omega_0$ & $R_0$ & $b$ & $M$ 
& $\left<\log|\delta_r|\right>$ & $\left<\log|\delta_\theta|\right>$ &
$T_0\,[\unit{MeV/k_B}]$\\
\hline
CR        & 3  & $5\!\times\!10^4$ & $10^5$ &  1.75 & 1.5 & $4\rho_n$ & 0  &
0.035 & $\infty$        & 0         & 2.17 & -7.0 & -7.6 & 0 \\
BC        & '' & ''            & ''   &  ''   & ''  & ''        & $2(\tilde \rho/\rho_0)^{5/8}$  &
''    & $\unit[15]{km}$ & 0         & 2.12 & -7.0 & -7.4 & 48\\
NC        & '' & ''            & ''   &  ''   & ''  & ''        & '' &
''    & ''              & $-2$      & 2.15 & -7.0 & -7.4 & '' \\
C$\Omega$ & '' & ''            & ''   &  ''   & ''  & ''        & '' &
''    & ''              & $-2^\ast$ & 2.16 & -4.2 & -4.1 & '' \\
C$p$      & '' & ''            & ''   &  ''   & ''  & ''        & '' &
''    & ''              & $-2^\ast$ & 2.15 & -4.6 & -5.6 & '' \\
BN        & '' & ''            & ''   &  ''   & ''  & ''        & $2-\tilde \rho/\rho_0$ & '' &
''    & 0               & 2.09 & -7.0 & -7.2 & 24 \\
NN        & '' & ''            & ''   &  ''   & ''  & ''        & '' &
''    & ''              & $-2$ & 2.12 & -6.9 & -7.1 & '' \\
\hline
\hline
\end{tabular}
\caption{Parameters and properties of the stellar models considered in this work.
The first column is the name of the model (see Sec.~\ref{ssec:models}),
columns 2--6 are the EOS
parameters, columns 7--11 are the parameters of the potential $Q$, and
columns 12--15 are model properties.
Symbol ` '' ' means ``same as above'' and the asterisk means that $b$ was included
in a non-consistent way in C$\Omega$ and C$p$.
See text for details.}
\label{tab:models}
\end{table*}

\subsection{Test 1: barotropic limit}

We checked that, using the non-barotropic inversion of the EOS (namely steps
5.a--5.b in Sec.~\ref{ssec:xns}), we obtain the same stationary results for
the cold, rigid rotating model CR (having drop the $\mathcal F$ term) and for the
barotropic, differentially rotating models BC and BN.

\subsection{Test 2: first integral residual}

We define the residuals of the Euler equation as
\begin{equation}
\label{eq:res}
\delta_i(r,\theta)= \partial_iQ(r,\theta) 
- \frac{\partial_ip(r,\theta)}{\hden(r,\theta)}
- F(r,\theta)\partial_i\Omega(r,\theta),
\end{equation}
where $i={r,\theta}$ is the direction of differentiation.
To quantify how well Eq.~\eqref{eq:iomes} is solved in the star we
use the averaged logarithm of the residuals:
\begin{equation}
\label{eq:avglogres}
\left<\log\left|\delta_i\right|\right>=\frac{\sum_j\log_{10}\left|\delta_i(r_j,\theta_j)\right|}N,
\end{equation}
where $j$ is the index that identifies a point inside the star and $N$
is the total number of points inside the star.
These quantities
should be compared with the potential $Q$ which is in the range
$0.3\lesssim Q\lesssim 0.8$.  We report the residuals in
Table~\ref{tab:models}. As expected, the Euler equation has in average a much
worse residual (2-3 orders of magnitude) in the control configurations
than in the consistently determined ones, thus corroborating our theory.

\subsection{Test 3: stellar oscillations}

\begin{figure}
\includegraphics[width=\columnwidth]{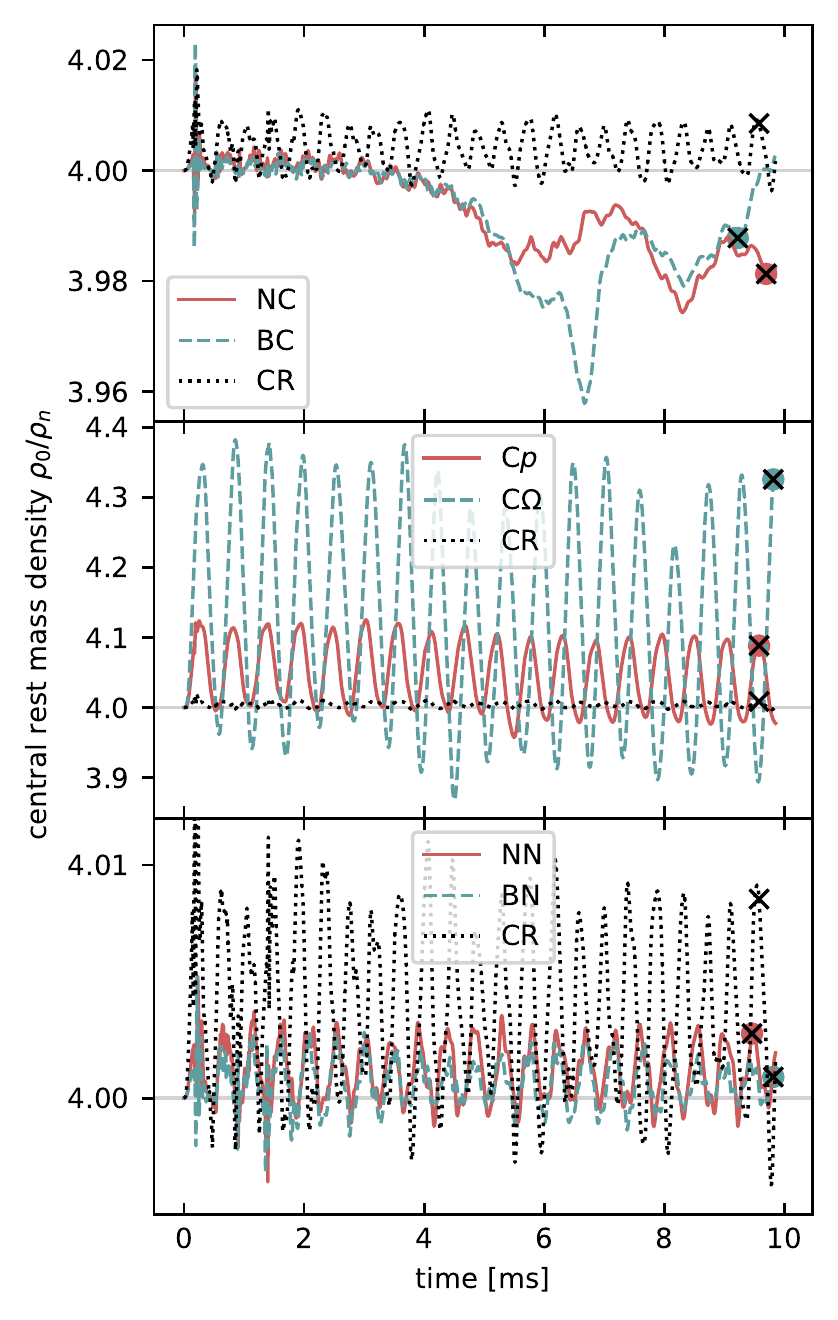}
\caption{\label{fig:osc}Time dependence of the central rest mass density in the BAM evolution
for the models considered in this paper. The cold, rigidly rotating model CR
is plotted in all panels as reference.  The crosses mark the snapshots shown in
Fig.~\ref{fig:bam} and the
gray horizontal lines mark the initial central density.}
\end{figure}

\begin{figure*}
\includegraphics[width=\textwidth]{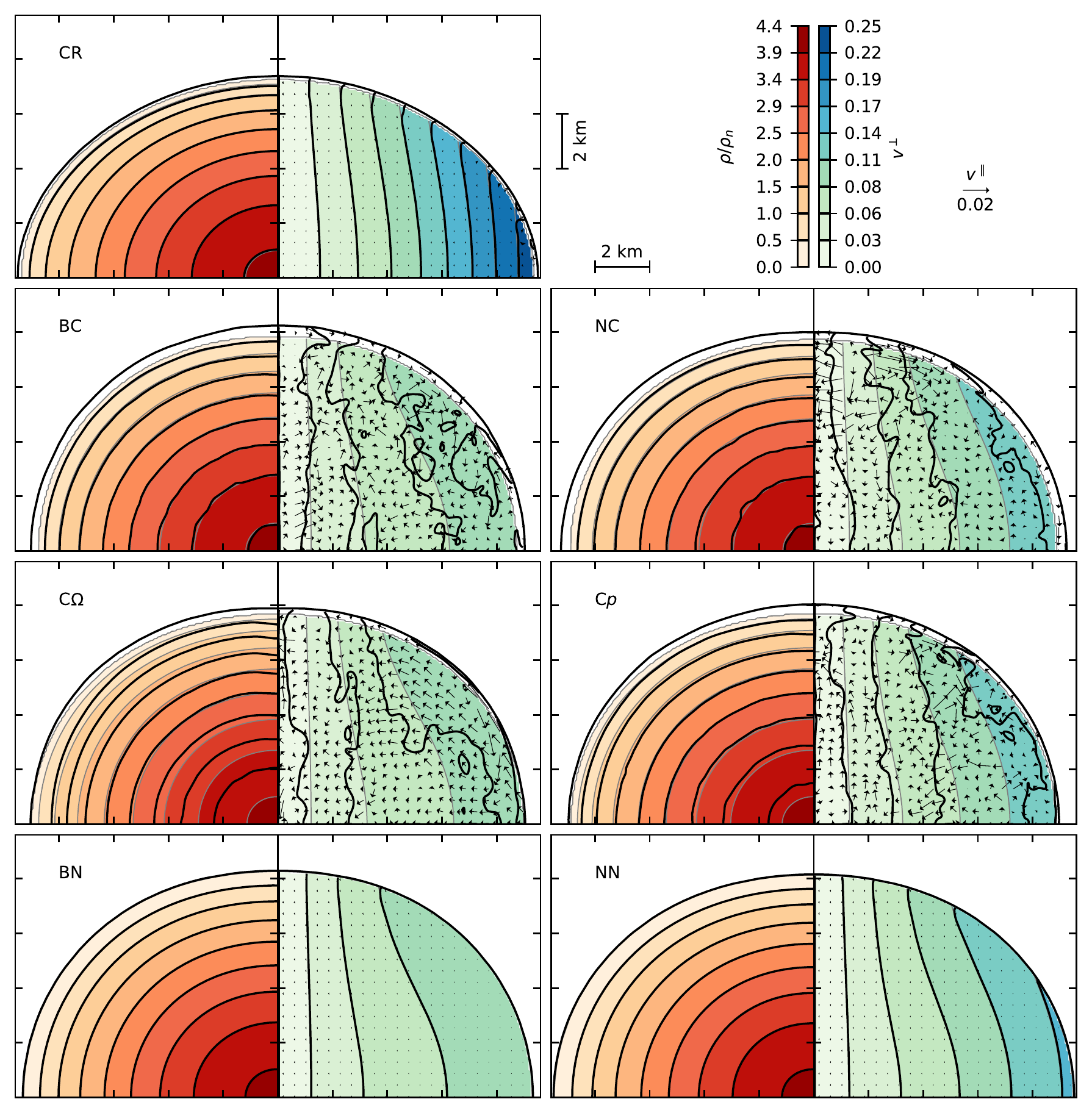}
\caption{\label{fig:bam}BAM evolution. For each model,
we plot the density $\rho$ (red scale, left)
and the orthogonal velocity $v^\perp=r\sin(\theta)v^\phi$ (blue scale, right).
The initial configurations are shown in color filled contours delimited by thin gray contours
while the configurations marked in Fig.~\ref{fig:osc} are shown in black thick contours
with the parallel velocity $v^\parallel=v^r\mathbf e_r + rv^\theta\mathbf
e_\theta$ shown as a vector field. Any deviance from stationarity during
the evolution is due to
convection and/or to the non consistency of the initial setup.
See text for details.}
\end{figure*}

As a final check, we evolved the XNS models with BAM to see whether
the configurations are indeed in equilibrium.
In particular, we want to compare the amplitude of the
oscillations that are artificially triggered by numerical inaccuracies
and by the use of the XCFC approximation for the initial setup.  In Fig.~\ref{fig:osc}
we show the central rest mass density evolution, and in Fig.~\ref{fig:bam} we
compare the initial configuration with a snapshot close to the maximum of the
final oscillation (marked with crosses in Fig.~\ref{fig:osc}), in such a way to maximize deviations.  Indeed,
control configurations diverge much more than the consistently
determined ones.

However, as discussed in Appendix~\ref{app:baro}, models BC and NC are
unstable against convection (note the convective patterns in the velocity field for these configurations in Fig.~\ref{fig:bam}).
Moreover, the convective timescale
is comparable with the evolution time (Appendix~\ref{app:baro}), and therefore
also these consistently determined stellar configurations deviate from the
initial ones.

We thus evolved 2
models that are stable against convection, BN
and NN. These configurations have small oscillations
comparable to that of the cold rigidly rotating model CR, thus verifying
our theory.

\begin{figure*}
\includegraphics[width=\textwidth]{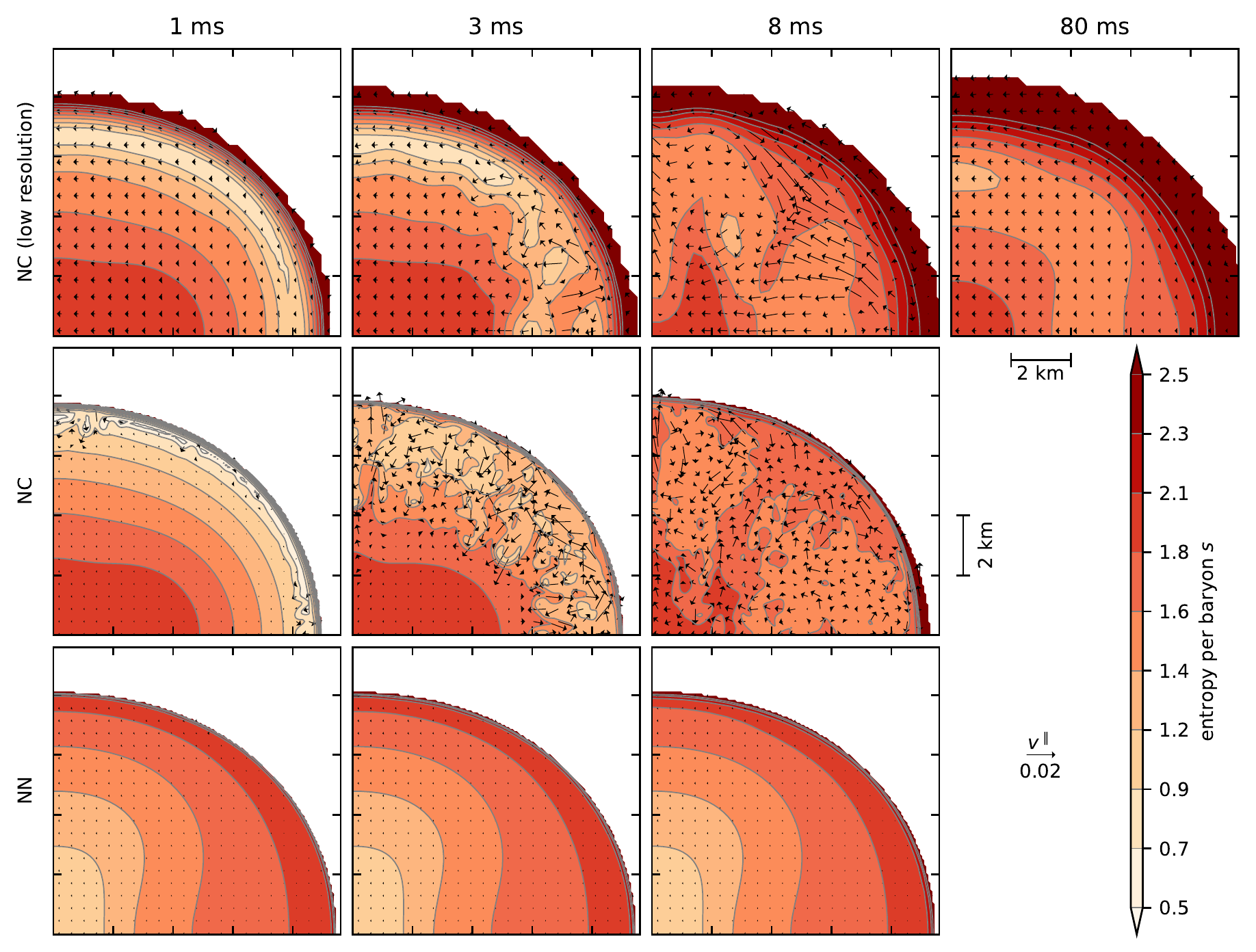}
\caption{\label{fig:con} Convection in the BAM evolution. Each row refers to a
different stellar model and each column to a different time snapshot.  The
entropy per baryon $s$ is shown as color filled contours and the parallel
velocity $v^\parallel=v^r\mathbf e_r + rv^\theta\mathbf e_\theta$
as a vector field. See text for details.}
\end{figure*}

In Fig.~\ref{fig:con} we compare the evolution of the non-barotropic setup
for the convective and non-convective star. Convection begins at the stellar
surface, where the convective timescale is shorter (Appendix~\ref{app:baro}),
and propagates to the interior, destroying the non-barotropic pattern and
flattening the entropy profile.
We have also simulated the evolution of a low resolution NC setup for a much longer
time. This low resolution simulation reproduces the
qualitative patterns of the high resolution one and in it the convective cells disappear
after $t\simeq\unit[10]{ms}$, in line with the qualitative estimates of the convective
timescale made in
Appendix~\ref{app:baro}\footnote{We note the larger entropy at the star's surface for the low resolution NC model. 
This entropy production is caused by the surface as discussed, e.g., in~\citet{Guercilena:2016fdl}. 
The entropy production decreases with an increasing resolution and its origin lies in the high-resolution 
shock-capturing schemes and the use of an artificial atmosphere surrounding the star.}.

As final remarks, we point out that:
\begin{itemize}
\item The control models too are unstable against convection;
however the non-consistency of the initial configurations has a much larger
destabilizing effect, cf.~Fig.~\ref{fig:osc}.
\item It is possible to obtain equilibrium models of neutron stars that are
unstable against convection as it is possible to obtain
equilibrium models that are dynamically unstable (i.e., that collapse \cite{Camelio+2018}).
\end{itemize}

\section{Discussion}
\label{sec:discussion}

\subsection{Consequences}

In the following we list some general results that can be directly derived with our novel approach:
\begin{enumerate}
\item The Schwarz's theorem implies that if $F=F(\Omega)$, then $s=s(p)$,
namely the EOS is an effective barotrope.  The vice versa is also true.
\item The Schwarz's theorem implies that a stationary neutron star with a non-barotropic thermal
profile must also be differentially rotating.
\item On the
symmetry axis $F$ vanishes; then if the star is barotropic [namely $\Omega=\Omega(F)$]
the angular velocity is uniform on the symmetry axis. However, this is not true
in general for a non-barotropic star (but it is for the non-barotropic cases considered in this
work) \cite{Clement1969, Monaghan1971, Uryu+Eriguchi1994,
Espinosa_Lara+Rieutord2007, Espinosa_Lara+Rieutord2013, Fujisawa2015}.
\item 
An interesting point that emerges from Sec.~\ref{sec:non-barotropic} is that
there are only two EOS quantities that can be directly determined from
the Euler equation without solving the EOS, namely $p$ and $\hden$.
This should not be a surprise because $p$ and $\hden$ are the only EOS quantities
that appear in the definition of the energy-momentum tensor, Eq.~\eqref{eq:tmunu}.
When other quantities like $\ln h$ and $s$ appear in the equations, they correspond
to physical quantities only in some limits, e.g., for isentropic stars in the
case of $\ln h$ and for barotropic stars in the case of $s$.
\item As already pointed out, the method we developed to obtain non-barotropic configurations
does not depend on the XCFC approximation and can be easily adapted to
the full stationary metric (even without the circularity assumption) or to Newtonian gravity (see Appendix~\ref{app:newton}).
All that is really needed to have non-barotropicity is that the potential
$Q$ depends on more than just the pressure [e.g., $Q(p,x)$], that the second free variable $x$
has a spatial distribution different from $p$, and that the cross partial derivative
of the potential $\partial_p\partial_x Q$ is not null. In this paper we chose the second variable
to be the angular velocity, $x = \Omega$, and we therefore consider differentially rotating neutron
stars, but in principle we could as well have used the magnetic field or meridional
currents \cite{Birkl+2011} instead (or in addition).
\item It is known that the numerical solution of the Euler equation for a
Newtonian non-barotropic star shows a degeneracy in the profile of $\Omega$
that can be lift by e.g.\@ including viscosity
\cite{Espinosa_Lara+Rieutord2013}.  This degeneracy does not arise in our
method because we fix the potential $Q(p,\Omega)$
and therefore we implicitly fix the profile of $\Omega$.
\end{enumerate}
Note that points 1 and 2 are a reformulation of the relativistic von Zeipel's theorem
\cite{Zeipel1924, Abramowicz1971}.

\subsection{General entropy profile}
\label{ssec:s}

In principle, it is possible to use the formalism
developed in this paper to determine the rotating profile of a
hot neutron star given its 2D thermal profile $s=s(r,\theta)$.

Let us assume a potential that further generalizes $Q(p,\Omega)$ in
Eq.~\eqref{eq:case_c}, for example
\begin{equation}
\label{eq:i_par}
Q(p,\Omega) = \sum_{l,m} a_{lm} H^l(p)\mathcal F^m(\Omega),
\end{equation}
where $a_{lm}$ are parameters and $H$ and $\mathcal F$ are formally defined as
before.
Now, given a choice of $a_{lm}$, we obtain a unique profile $s(r,\theta)$ from the
solution of Eqs.~\eqref{eq:sys1}--\eqref{eq:sys3}. To ensure that the entropy in a given
point within the star takes a specified value,
$s(r',\theta')=s'$, one can modify the potential free parameters, e.g.,
$a_{l'm'}$. If we want to fix the entropy in two points, we must tweak two
free parameters, and so on. In principle we can fix the entropy in all grid
points by adjusting an equal number of parameters.

In practice, the procedure described above may be cumbersome if one wants to fix
the entropy in more than a few points and we discussed it only as a proof of principle.
Moreover, this procedure works only for \emph{planar} configurations, namely
$s(r,\theta)=s(r,\pi-\theta)$.  To obtain a non-planar configuration one should
define two potentials $Q_-$ and $Q_+$ that coincide together with their first
and second partial derivatives along a given curve $\big(p(z),\Omega(z)\big)$,
where $z$ is the curve parameter.

We remark that this procedure would work also if one wants
to fix the rotational profile $\Omega=\Omega(r,\theta)$ instead of the
entropy one.

\subsection{Multi-dimensional equation of state}
\label{ssec:eos3d}

Let us consider an EOS that depends
on $N>2$ independent variables, e.g.\@ $h=h(p,s,Y)$, where $Y$
is the proton number fraction.

In this case one should solve Eqs.~\eqref{eq:sys1}--\eqref{eq:sys2} as for the
non-barotropic case of the EOS with two independent variables. The difference
is that Eq.~\eqref{eq:sys3} now becomes
\begin{equation}
\label{eq:sys3bis}
\partial_p Q(p,\Omega)= \frac{1}{\hden\big(p,s(r,\theta),Y(r,\theta)\big)}.
\end{equation}
At this point, one can fix $Y(r,\theta)$
and invert the EOS
to determine $s(r,\theta)$. Another way to look at this is that
the 3D EOS is equivalent to a parameterized 2D EOS:
$\hden\big(p,s,Y(r,\theta)\big)=\hden_{Y(r,\theta)}(p,s)$.

We remark that:
\begin{itemize}
\item It is possible to fix $s(r,\theta)$ instead of $Y(r,\theta)$, but
not both profiles at the same time, unless one uses the
procedure discussed in Sec.~\ref{ssec:s}.
\item The results discussed above would stay valid when $s$ and/or $Y$ do not
explicitly depend on $(r,\theta)$ but on $(p,\Omega)$,
since all these quantities are known when one solves Eq.~\eqref{eq:sys3bis}.
\end{itemize}

\subsection{Legendre transformation}
\label{ssec:leg}

In thermodynamics, different choices of free variables imply the use of different
thermodynamical potentials, that are related to each other
by Legendre transformations.
What if we take the Legendre transformation of the potential $Q$?

First, we define the following transformed potential
\begin{equation}
\label{eq:leg_q}
\mathcal Q(p,F) = Q\big(p,\Omega(p,F)\big) - \Omega(p,F) F,
\end{equation}
where the independent variables are $p,F$ and therefore the angular velocity
is written as $\Omega=\Omega(p,F)$, cf.~Eq.~\eqref{eq:leg_h}.
The differential of Eq.~\eqref{eq:leg_q} yields
\begin{align}
\mathrm d\mathcal Q={}& \frac{\mathrm dp}{\hden} - \Omega \mathrm d F,\\
\hden^{-1}={}& \left.\frac{\partial\mathcal Q}{\partial p}\right|_F,\\
\Omega={}& -\left.\frac{\partial \mathcal Q}{\partial F}\right|_p,
\end{align}
where all quantities depend on $(p,F)$.

In order to re-obtain the barotropic, differentially rotating model
we assume that the EOS is an effective barotrope and that $\Omega=\Omega(F)$.
Similarly to what was done in Sec.~\ref{ssec:baro-diff}, we can define a function
$\mathcal G=\mathcal G(F)$ such that
\begin{equation}
\label{eq:der_o}
\Omega(F)= -\frac{\mathrm d \mathcal G(F)}{\mathrm dF}.
\end{equation}
The j-const differential-rotation law is equivalent to
\begin{equation}
\label{eq:o_law}
\mathcal G(F)= \left(\frac{\sigma^2}{2}F-\Omega_0\right)F,
\end{equation}
where $\sigma=1/R_0$ is a parameter.
The barotropic potential of Eq.~\eqref{eq:case_c} is equivalent to the following barotropic transformed potential:
\begin{equation}
\mathcal Q(p,F)= H(p) + \mathcal G (F) - \ln\alpha_0.
\end{equation}
An advantage of this formulation is that it unifies rigidly and differentially
rotating stars.  Indeed, the rigid rotation limit $R_0\to\infty$ corresponds to
$\sigma=0$ and therefore $\Omega(F)\equiv \Omega_0$ is well defined. It 
also simplifies the inclusion of differential rotation laws where $F(\Omega)$ is
not monotonic \cite{Uryu+2017}, which are a more realistic description of post-merged neutron stars.

\section{Conclusions}
\label{sec:conclusions}

In this paper we have studied, for the first time, a stationary, differentially rotating,
non-barotropic neutron star
in General Relativity.  In doing so, we have shown with theoretical
arguments and with stationary and dynamical numerical simulations how the
Euler equation can be cast in a potential form also in the non-barotropic
case.
This is a novel results even in the context of Newtonian stars.

To test our approach, we have first generated stationary configurations using
the XNS code \cite{Bucciantini+Del_Zanna2011,
Pili+2014, Camelio+2018}, that determines the neutron star
structure and spacetime in the eXtended Conformal Flatness
Condition approximation \cite{Cordero-Carrion+2009}. 
We have then taken the stationary configurations as initial
condition for dynamical evolutions performed with 
the general relativistic hydrodynamics code
BAM~\cite{Brugmann:2008zz,Thierfelder:2011yi}.  We considered consistently
determined configurations of barotropic and non-barotropic rotating neutron
stars and compared them with non-consistent ``control'' configuration to gauge the quality
of our models.  We considered both convectively stable
and unstable models.

We used our formalism to demonstrate some properties of non-barotropic
stars, most notably that a non-barotropic star must be differentially rotating
\cite{Zeipel1924, Abramowicz1971}
and that in a non-barotropic star the specific angular momentum and the entropy must
depend on both pressure and angular velocity.\\

Possible outlooks of this work are the following.

One can use the final snapshots of dynamical evolutions to model the
Euler equation potential of (i) post merged neutron stars, (ii) proto neutron stars (post core collapse),
and (iii) post hadron phase transition quark stars. Then, one can quickly explore the
parameter space of the hot rotating remnant with a stationary code like XNS
to study e.g.\@ the dynamical stability, the maximal mass, the gravitational wave signal from stellar
quasi-periodic oscillations, etc.
The most interesting configurations can then be selected to be further explored with
dynamical codes like BAM, using the XNS output as completely consistent initial data \cite[e.g.,][]{Camelio+2018}.

In Sec.~\ref{ssec:s} we showed how in principle is possible to use our
potential formalism to determine a general entropy profile. 
But another, maybe simpler, method would be to import the techniques developed in the
context of Newtonian baroclinic stars to include a general thermal profile.
In this way one can study the long term (on the order of minutes), neutrino-driven,
quasi-stationary evolution of the hot and rotating remnant of cases (i--iii) \cite{Burrows+Lattimer1986,
Keil+Janka1995, Pons+1999, Villain+2004, Roberts2012, Camelio+2016}.
This is important because a huge amount of energy (up to tenths of solar masses)
is expected to be radiated through neutrinos in the first phase of the neutron star life.
However, this phase is too long to be fully explored with dynamical codes, while
using a quasi-stationary evolution would allow to employ stationary, fast codes like XNS.
Again, in this way one can quickly study the parameter space and select
the most interesting configurations to be further explored with dynamical
codes, and even study the time dependent gravitational wave signal from
this phase \cite{Ferrari+2003, Camelio+2017} and assess the role of physical processes such as
viscosity.

Finally, one can apply our potential formalism to the study of non-barotropicity in accretion disks
\cite{Amendt+1989, Razdoburdin2017, Witzany+Jefremov2018}, in neutron stars with
magnetic field \cite{Chatterjee+2015} and with meridional currents
\cite{Birkl+2011}, and in Newtonian stars.

\appendix

\section{Newtonian limit}
\label{app:newton}

In the Newtonian limit,
\begin{align}
\alpha\to{}& \exp\Phi,\\
\hden\to{}&\rho,\\
R\to{}&\varpi=r\sin\theta,\\
v^\phi\to{}&\Omega,\\
F\to{}& j=\varpi^2\Omega,\\
\el\to{}& j=\varpi^2\Omega,\\
Q\to{}&-\left(\Phi - \frac12\varpi^2\Omega^2\right), 
\end{align}
where $\Phi$ is the gravitational potential, $\varpi$ the cylindrical radius,
and $j$ the non-relativistic specific angular momentum.  Note that both $F$
and $\el$ tend to the same limit: $j$, and that the potential $Q$ [Eq.~\eqref{eq:i_lnaw}] tends to
minus the effective (including the centrifugal force) gravitational potential.
The Newtonian limit of Eq.~\eqref{eq:iomes} is:
\begin{equation}
\label{eq:euler}
\frac{\nabla p}\rho + \nabla\left(\Phi - \frac12\varpi^2\Omega^2\right)
+ j\nabla\Omega=0,
\end{equation}
where $i=\{r,\theta\}$ and we divided by $r$ the equation along the $\theta$ direction.

Eq.~\eqref{eq:euler} is equivalent to the stationary Euler equation
adopted in the Newtonian literature,
[e.g., Eqs.~(2)--(3) of
Ref.\@ \onlinecite{Uryu+Eriguchi1994} and Eq.~(20) of Ref.\@
\onlinecite{Espinosa_Lara+Rieutord2013}]
\begin{equation}
\label{eq:euler-literature}
\frac{\nabla p}\rho + \nabla\Phi - \varpi\Omega^2\mathbf{e}_\varpi=0,
\end{equation}
where $\mathbf{e}_\varpi$ is a unit vector along the cylindrical radius
and we assumed circular motion (i.e., no meridional currents) and no viscosity.

We will show here that Eqs.~\eqref{eq:euler}--\eqref{eq:euler-literature} are equivalent
by recovering both from the general form of the stationary (Newtonian) Euler equation:
\begin{equation}
(\mathbf{v}\cdot\nabla)\mathbf{v}= -\frac{\nabla p}{\rho} - \nabla\Phi,
\end{equation}
where $\mathbf{v}=\Omega\varpi\mathbf{e}_\phi$ is the fluid velocity.
From the identity
\begin{equation}
(\mathbf{v}\cdot\nabla)\mathbf{v}=\frac12\nabla (\mathbf{v}\cdot\mathbf{v}) - \mathbf{v}\times (\nabla\times\mathbf{v}),
\end{equation}
we get
\begin{align}
(\mathbf{v}\cdot\nabla)\mathbf{v}={}&
\frac12\nabla(\varpi^2\Omega^2)-x,\\
\label{eq:x}
x={}&\Omega\sin\theta\partial_r(r\varpi\Omega) \mathbf{e}_r + \Omega\partial_\theta(\sin\theta\varpi\Omega)\mathbf{e}_\theta.
\end{align}
Now, if we directly expand the partial derivatives in $x$,
\begin{multline}
x= \left(\mathbf{e}_r\partial_r+\frac{\mathbf{e}_\theta}{r}\partial_\theta\right)\frac{\varpi^2\Omega^2}2\\
+ \varpi\Omega^2(\sin\theta\mathbf{e}_r + \cos\theta\mathbf{e}_\theta),
\end{multline}
we recover Eq.~\eqref{eq:euler-literature}.

On the other hand, we have also
\begin{multline}
x=\Omega\partial_r(\varpi^2\Omega)\mathbf{e}_r + \frac\Omega{r} \partial_\theta(\varpi^2\Omega)\mathbf{e}_\theta\\
=\nabla(\varpi^2\Omega^2) - \varpi^2\Omega\nabla\Omega,
\end{multline}
from which we recover Eq.~\eqref{eq:euler}.

Our non-barotropic potential formalism can be simply extended to the Newtonian
case by applying it to Eq.~\eqref{eq:euler}.

\section{2D equation of state}
\label{app:eos}

We choose a polytropic expression for the total energy per baryon:
\begin{align}
e(\rho,s) ={}&m_n\big(1+ u_\mathrm{cold}(\rho) + u_\rth(\rho,s)\big),\\
u_\mathrm{cold}(\rho)={}&k_1 \rho^{\Gamma - 1},\\
u_\rth(\rho,s)={}&k_2 s^2 \rho^{\Gamma_\mathrm{th}-1},
\end{align}
where $\rho$ is the rest mass density, $s$ the entropy per baryon, $m_n$ is the nucleon mass,
$u_\mathrm{cold}$ the specific cold internal energy, $u_\rth$ the
specific thermal internal energy, and $k_1,\Gamma,k_2$ and $\Gamma_\rth$ are parameters.
We remark that to have physical results for any physical $\rho,s$ it has to be $\Gamma>1,\Gamma_\mathrm{th}>1,k_1>0$, and $k_2\ge 0$.
Using the relation (that is a consequence of the first law of Thermodynamics)
\begin{equation}
\frac p{\rho^2}=\frac1{m_n}\left.\frac{\partial e}{\partial \rho}\right|_s,
\end{equation}
where $p$ is the pressure, we get 
\begin{equation}
\label{eq:prhos}
p(\rho,s)=(\Gamma - 1)\rho u_\mathrm{cold}(\rho) + (\Gamma_\mathrm{th} - 1)\rho u_\rth(\rho,s).
\end{equation}
Eq.~\eqref{eq:prhos} can be written as
\begin{align}
p(\rho,u_\rth)={}& K\rho^\Gamma + (\Gamma_\mathrm{th} - 1)\rho u_\rth,\\
K={}&(\Gamma - 1)k_1,
\end{align}
namely we recover Eq.~\eqref{eq:ideal_gas}.

Using the thermodynamical relation
\begin{equation}
T=\left.\frac{\partial e}{\partial s}\right|_\rho,
\end{equation}
where $T$ is the temperature, we obtain
\begin{equation}
T(\rho,s)=2m_nk_2s\rho^{\Gamma_\rth -1}.
\end{equation}
We remark that $T\to 0$ as $s\to 0$, as expected.

The speed of sound is defined by
\begin{equation}
\label{eq:cs}
c_s= \sqrt{\left.\frac{\partial p}{\partial \epsilon}\right|_s},
\end{equation}
where $\epsilon=\rho e/m_n$ is the total energy density.
For our EOS it is
\begin{equation}
\label{eq:cs2}
c_s^2= \frac{\Gamma(\Gamma-1)k_1\rho^{\Gamma} + \Gamma_\rth(\Gamma_\rth-1)s^2k_2\rho^{\Gamma_\rth}}
{\rho + \Gamma k_1 \rho^{\Gamma} + \Gamma_\rth s^2 k_2\rho^{\Gamma_\rth}}.
\end{equation}

From the Legendre transformation of the specific energy
\begin{equation}
\label{eq:leg_h}
h(p,s)=\frac{e\big(\rho(p,s),s\big)}{m_n} + \frac{p}{\rho(p,s)},
\end{equation}
we get the specific enthalpy $h$
\begin{multline}
h(p,s) = 1 + \Gamma k_1 \big(\rho(p,s)\big)^{\Gamma-1}\\ + \Gamma_\mathrm{th}
k_2 s^2 \big(\rho(p,s)\big)^{\Gamma_\mathrm{th} - 1},
\end{multline}
where $\rho(p,s)$ is the inverse of Eq.~\eqref{eq:prhos}. The reason why we write all quantities
in terms of $p$ and $s$ is that $h$ is naturally a function of these variables, see discussion in Sec.~\ref{ssec:cr}.

From the solution of the Euler equation [Eqs.~\eqref{eq:sys1}-\eqref{eq:sys3}]
we obtain $\hden$ and $p$, from which we want to get all the other
thermodynamical quantities.  To invert the EOS, we first cancel out the term with the entropy and obtain the equation
\begin{equation}
\label{eq:hden_p}
(\Gamma_\rth - 1) \hden - \Gamma_\rth p = (\Gamma_\rth - 1) \rho + (\Gamma_\rth - \Gamma)k_1\rho^\Gamma,
\end{equation}
where $\hden=h\rho$ is the enthalpy density and the only unknown is the density $\rho$.
This equation can be easily solved if $\Gamma=3/2$ (when it becomes cubic in $\sqrt{\rho}$),
$\Gamma=2$ (quadratic in $\rho$) or $\Gamma=3$ (cubic in $\rho$).

We pick $\Gamma=3$ because it is closer to the stiffness expected for the high-density part of the
real EOS \cite{Rosswog+Davies2002}.
We can at this point set the parameter $k_1$ enforcing the condition
$2.1\lesssim M_\mathrm{max}\lesssim 3$, where $M_\mathrm{max}$ is the maximal non-rotating mass.

We choose $\Gamma_\rth=1.75$, which is a value that reproduces
the behavior of known finite-temperature EOSs \cite{Bauswein+2010, Yasin+2018}.
To set $k_2$ we require that the thermal contribution to the pressure at $\rho=2\rho_n$ and
$s=\unit[2]{k_B}$ is approximately $30\%$, value determined by inspection of realistic EOSs.
The corresponding temperature is
$T(2\rho_n,\unit[2]{k_B})\simeq\unit[29]{MeV/k_B}$.

The solution of Eq.~\eqref{eq:hden_p} is not always unique.
In particular, when $\Gamma_\rth<\Gamma=3$ there
are values of $(\hden,p)$ which correspond to two valid solutions
$(\rho_1,s_1)$ and $(\rho_2,s_2)$ with
$\rho_1\le\rho_c\le\rho_2$, where $\rho_c$ is a critical
density that depends on $\Gamma,\Gamma_\rth,k_1$:
\begin{equation}
\label{eq:rhocritical}
\rho_c=\sqrt{\frac{\Gamma_\rth - 1}{3k_1(\Gamma-\Gamma_\rth)}}\qquad(\Gamma_\rth<\Gamma=3).
\end{equation}
A way around this difficulty is to choose a stellar configuration such that the maximal density
is lower than $\rho_c$, in order to safely take the root $\rho_1$.

We report the EOS parameters in Table~\ref{tab:models}. With those, we get the
following EOS properties (cf.~Fig.~\ref{fig:mrho}):
\begin{itemize}
\item Critical density for EOS inversion: $\rho_c=4.61\rho_n$.
\item The speed of sound of the cold EOS becomes greater than the speed of
light at $\rho_{c_s}=5.95\rho_n$.
\item Central density of the (cold, non-rotating) maximal mass configuration:
$\rho_\mathrm{max}=6.90\rho_n$.
\item Maximal mass of the cold, non-rotating star: $M_\mathrm{max}=\unit[2.22]{M_\odot}$,
\end{itemize}
where $\rho_\mathrm{max}$ and $M_\mathrm{max}$ are obtained enforcing causality at densities greater than
$\rho_{c_s}$ and without attaching a crust at low densities.

Since all considered models have a central density $\rho_0=4\rho_n$ (see Table~\ref{tab:models}),
we avoid the problems related to causality and uniqueness.
This value is also smaller than the central density $\rho_\mathrm{max}$ of the
non-rotating maximal mass configuration; and since additionally we chose
$\Omega_0$ such that the gravitational (Komar) mass is smaller than (but close to) the maximal
non-rotating mass, then all studied models are dynamically stable (i.e., they do not collapse).

\section{Barotropic EOS}
\label{app:baro}

When the EOS is an effective barotrope every thermodynamical quantity depends only
on the pressure, for example $s=\tilde s(p), \rho=\tilde \rho(p), h=\tilde h(p), \ldots$
(we mark the barotropic functions with a tilde to stress that they correspond to physical
quantities only in a barotropic stellar model, while the pressure $p$ is always equivalent
to the physical quantity).

The easiest choice for the barotropic function is
\begin{equation}
\label{eq:s_rho_baro1}
\tilde s(p)= k_3 \big(\tilde \rho(p)\big)^\frac{\Gamma - \Gamma_\rth}2,
\end{equation}
where $k_3$ is a constant;
in this case the heat integral can be easily integrated
in $\tilde\rho$:
\begin{equation}
\label{eq:heat_rho}
H(p)=\int_{\tilde\rho(p_0)}^{\tilde\rho(p)}
\frac{p'(\tilde\rho)}
{\tilde\hden\big(p(\tilde\rho)\big)}\mathrm d\tilde\rho,
\end{equation}
where $p_0$ is the central pressure, $\tilde\hden$ is
the enthalpy density, $p(\tilde\rho)$ is the inverse of $\tilde\rho(p)$, and $p'(\tilde\rho)$ is
its total derivative with respect to $\tilde\rho$.
Indeed, in this case we analytically obtain
\begin{equation}
H(p)= \frac{\Gamma[(\Gamma-1)k_1 + (\Gamma_\mathrm{th}-1)k_2k_3^2]}
{(\Gamma-1)[\Gamma k_1 + \Gamma_\mathrm{th}k_2k_3^2]}\ln \frac{\tilde h(p)}{\tilde h_0}.
\end{equation}
where $\tilde h_0$ is the central specific enthalpy.
We remark that $H=\ln h/h_0$ when $k_2k_3^2=0$ (i.e., cold star) or $\Gamma=\Gamma_\rth$
(i.e., isentropic star), as it should be.

We will consider another possibility for the barotropic function:
\begin{equation}
\label{eq:s_rho_baro2}
\tilde s(p)= \tilde s_s - k_3\tilde \rho(p),
\end{equation}
where $\tilde s_s$ is the surface entropy and $k_3$ a constant.
Unfortunately in this case there is no simple analytical form for
the heat integral and we integrate Eq.~\eqref{eq:heat_rho}
numerically.

Let us now consider the stability of the star against convection.  We will use
the convective criterion in spherical symmetry, namely for a non-rotating
neutron star, as an estimate for our rotating case.
In spherical symmetry the star is unstable against convection
when the Schwarzschild discriminant is negative \cite{Thorne1966},
\begin{equation}
\label{eq:S_discriminant}
S(\bar r)=\frac{\mathrm dp}{\mathrm d \bar r}
- c_s^2 \frac{\mathrm d\epsilon}{\mathrm d \bar r} <0,
\end{equation}
where $c_s$ is the speed of sound [Eq.~\eqref{eq:cs2}] and the total derivatives
are taken along the Schwarzschild radius $\bar r$ that is related to the isotropic radius by
\begin{equation}
\label{eq:rbar_r}
\frac{\mathrm d\bar r}{\sqrt{\bar r^2 -2 m(\bar r)\bar r}} = \frac{\mathrm dr}r,
\end{equation}
where $m(\bar r)$ is the gravitational mass enclosed in $\bar r$.

For our EOS, Eq.~\eqref{eq:S_discriminant} is equivalent to
\begin{equation}
\label{eq:cond_conv}
\left[(\Gamma_\rth-1) + k_1\Gamma(\Gamma_\rth - \Gamma)\rho^{\Gamma-1}\right]
\frac{\mathrm ds}{\mathrm d\bar r}<0.
\end{equation}
For our choice of $\Gamma_\rth < \Gamma$, this means that if the entropy
gradient is negative (resp.~positive) there is convection when $\rho<\rho_c$ (resp.~$\rho>\rho_c$),
where the critical density for convection $\rho_c$ happens to be equal to the critical
density for inverting the EOS, Eq.~\eqref{eq:rhocritical}.
Then, since in our models the rest mass density is always smaller than $\rho_c$,
we expect convection for
barotropic profiles given by Eq.~\eqref{eq:s_rho_baro1} and vice versa no convection
for barotropic profiles given by Eq.~\eqref{eq:s_rho_baro2}.

For the case with convection, the convective timescale is given by
the analytical estimate \cite{Thorne1966}
($g$ is the strength of the gravity acceleration)
\begin{equation}
\label{eq:conv_tau}
\tau_c= c_s\sqrt{\frac{2\hden}{-gS(\bar r)}},
\end{equation}
which is of the order of tens of milliseconds close to the stellar center
and reduces to a timescale of the order of $\unit[0.1]{ms}$ close to the stellar surface
(these timescales are compatible with those found by \citet{De_Pietri+2018}
and \citet{De_Pietri+2019} in their simulations).
This means that we expect convection to influence our dynamical
simulations (that last for 10~ms), and that it starts
at the surface and propagates to the center.

While this analysis is strictly valid only for a non-rotating barotropic star,
we find that its application to rotating non-barotropic stars
qualitatively agrees with the results obtained from dynamical simulations
[Sec.~\ref{sec:results}].

\begin{acknowledgments}
TD acknowledges support by the European Union's Horizon 2020 research and
innovation program under grant agreement No 749145, BNSmergers.  SR has been
supported by the Swedish Research Council (VR) under grant number 2016-03657 3,
by the Swedish National Space Board under grant number Dnr. 107/16 and by the
VR research environment grant “Gravitational Radiation and Electromagnetic
Astrophysical Transients” (GREAT) under  Dnr. 2016-06012.  Support from the
COST Actions on neutron stars (PHAROS; CA16214) and black holes and
gravitational waves (GWerse; CA16104) are gratefully acknowledged.

We are grateful to M.A.~Abramowicz, B.~Br\"ugmann, S.~Faraji, C.~Lundman,
J.A.~Pons, and M.~Rieutord for useful discussions and comments on the paper
draft.
\end{acknowledgments}

\bibliography{paper.bbl}

\end{document}